\documentclass[pdftex,twocolumn,epjc3]{svjour3}          

\RequirePackage[T1]{fontenc}

\smartqed  

\RequirePackage{graphicx}
\RequirePackage{mathptmx}      
\RequirePackage{flushend}
\RequirePackage[numbers,sort&compress]{natbib}
\RequirePackage[colorlinks,citecolor=blue,urlcolor=blue,linkcolor=blue]{hyperref}
\usepackage{amsmath}
\journalname{Eur. Phys. J. C}

\begin{document}

\title{Minimal Spin-3/2 Dark Matter in a simple $s$-channel model}

\author{Mohammed Omer Khojali\thanksref{e1}, Ashok Goyal\thanksref{e2}\thanksref{t1}, Mukesh Kumar\thanksref{e3} 
	\and
	Alan S. Cornell\thanksref{e4} 
}

\thankstext[$\star$]{t1}{Visiting researcher}
\thankstext{e1}{e-mail: khogali11@gmail.com}
\thankstext{e2}{e-mail: agoyal45@yaho.com}
\thankstext{e3}{e-mail: mukesh.kumar@cern.ch}
\thankstext{e4}{e-mail: Alan.Cornell@wits.ac.za}
\institute{National Institute for Theoretical Physics,
School of Physics and Mandelstam Institute for Theoretical Physics,
University of the Witwatersrand, Johannesburg,
Wits 2050, South Africa.}

\date{Received: date / Accepted: date}

\maketitle

\begin{abstract}
We consider a spin~-~3/2 fermionic dark matter candidate (DM) interacting with Standard Model fermions through
a vector mediator in the $s$-channel. We find that for pure vector 
couplings almost the entire parameter space of the DM and mediator mass consistent with the observed 
relic density is ruled out by the direct detection observations through DM-nucleon elastic scattering 
cross-sections. In contrast, for pure axial-vector coupling, the most stringent constraints are 
obtained from mono-jet searches at the Large Hadron Collider.  
\end{abstract}

\section{Introduction}
\label{sec:intro}
A large number of cosmological and astrophysical observations provide strong evidence for the existence 
of Dark Matter (DM) in the universe. The amount of cold dark matter (CDM) has been precisely estimated 
from the measurements of the Planck satellite to be 
$\Omega_{\text{DM}} h^{2} = 0.1198 \pm 0.0015$~\cite{Ade:2015xua}. 
The nature of DM particles and their properties is the subject of intense investigation. One of the main physics 
programmes at the Large Hadron Collider (LHC) is devoted to the detection of DM, where there is the real 
possibility of the production of DM particles of any spin at 13 TeV centre-of-mass energy. As such, the ATLAS and 
the CMS collaborations are closely examining several DM signatures involving missing energy, \({\not}E_{T}\), 
accompanied by a single or two jet events~\cite{Askew:2014kqa}. In addition there are direct detection 
experiments, which measure the nuclear-recoil energy and its spectrum in DM-nucleon elastic scattering. The 
indirect detection experiments look for signals of DM annihilation into Standard Model (SM) particles in cosmic 
rays, and have detection instruments mounted on satellites and ground based 
telescopes~\cite{Schumann:2013mya, Cirelli:2015oqy}.      

Effective field theories (EFT) in which the DM-SM interactions are mediated by heavy particles, which are not 
accessible at the LHC energies, have been analysed in detail with limits from direct and indirect searches. 
Recently the need to go beyond these EFT models has been pointed out, in light of the large energy 
accessible at the LHC~\cite{Abdallah:2014hon}. Simplified models of DM with interactions to SM particles 
have emerged as attractive alternatives to EFT models. In these models the interaction between the DM and SM 
particles are mediated by spin-0 and spin-1 particles in the $s$-channel, whereas in the $t$-channel models the 
mediator can be a scalar, a fermion or a vector particle which will typically also carry colour or lepton 
number~\cite{Abdallah:2014hon}.

In this paper we consider a minimal SM singlet spin~-~3/2 vector-like fermion, \(\chi\), as a DM candidate, 
interacting with the SM particles through the exchange of a spin-1 mediator, \(Z^{\prime}\), in a minimal flavour 
violation (MFV) $s$-channel model. Spin-3/2 particles exist in several models beyond the SM, namely in models 
of supergravity where the graviton is accompanied by spin~-~3/2 gravitino superpartner. Spin-3/2 fermions also 
exist in Kaluza-Klein models, in string theory, and in models of composite 
fermions~\cite{C:JBH, ArkaniHamed:1998rs, L:RSP}. 
Recently spin~-~3/2 CDM has been studied  in EFT models, and constraints from direct and indirect observations 
obtained~\cite{Yu:2011by,Ding:2012sm,Ding:2013nvx,Savvidy:2012qa}. Spin-3/2, 7.1 keV Warm Dark Matter (WDM) has 
also been considered as a means to provide a viable explanation from the anomalous 3.1 KeV X-ray line observed 
by the XMM Newton~\cite{Dutta:2015ega}. As such we shall introduce the spin~-~3/2 CDM in an MFV 
$s$-channel model in section~\ref{sec:2}. Whilst in section~\ref{sec:3} we discuss all relevant experimental 
constraints including the relic 
density and the signatures of these DM particles at the LHC. In section~\ref{sec:4} we summarise 
our main results.
   
\section{Spin-3/2 Singlet DM Model}
\label{sec:2}
In this paper we extend the SM by including a spin~-~3/2 particle \(\chi\). We further let \(\chi\) to 
be a SM singlet which interacts with the SM particles through the exchange of a vector particle \(Z^{\prime}\) 
in the $s$-channel. Note that this can be done, for example, by extending the SM gauge symmetry with a 
new $U(1)^{\prime}$ gauge symmetry which is spontaneously broken, such that the mediator obtains a mass 
$m_{Z^{\prime}}$. We also invoke a discrete ${Z}_{2}$ symmetry under which the spin~-~3/2 DM particle \(\chi\) 
is odd, whereas all other SM particles, including the vector mediator \( Z^{\prime}\), are even. The spin~-~3/2 free 
Lagrangian is given by~\cite{Christensen:2013aua}: 
\begin{equation}
\mathcal{L} = \bar{\chi}_{\mu} \Lambda^{\mu\nu} \chi_{\nu},
\end{equation}
with
\begin{equation}
\Lambda^{\mu\nu} = (i{\not}\partial-m_{\chi})g^{\mu\nu} - i(\gamma^{\mu}\partial^{\nu} + \gamma^{\nu} \partial^{\mu})
 + i\gamma^{\mu}{\not}\partial\gamma^{\nu} + m_{\chi}\gamma^{\mu}\gamma^{\nu}.
\end{equation}
Note that \(\chi_{\mu}\) satisfies \(\Lambda^{\mu\nu}\chi_{\nu} = 0\), and with $\chi_\mu$ being on mass-shell we have 
\begin{equation}
(i{\not}\partial-m_{\chi})\chi_{\mu} = \partial^{\mu} \chi_{\mu} = \gamma^{\mu}\chi_{\mu} = 0.
\end{equation} 
The spin sum for spin~-~3/2 fermions 
\begin{equation}
S_{\mu\nu}^{+}(p) = \sum_{i = -3/2}^{3/2} u^{i}_{\mu}(p)\bar{u}^{i}_{\nu}(p)
\end{equation}
and 
\begin{equation}
S_{\mu\nu}^{-}(p) = \sum_{i = -3/2}^{3/2} v^{i}_{\mu}(p)\bar{v}^{i}_{\nu}(p),
\end{equation}
are given by~\cite{Christensen:2013aua}:
\begin{align}
S^{\pm}_{\mu\nu}(p) =& -({\not}p \pm m_{\chi}) \Big[g_{\mu\nu} - \frac{1}{3}\gamma_{\mu}\gamma_{\nu} 
-\frac{2}{3m^{2}_{\chi}}{\not} p_{\mu}{\not} p_{\nu}
\nonumber\\
&\mp\frac{1}{3m_{\chi}}(\gamma_{\mu}p_{\nu} - \gamma_{\nu}p_{\mu})\Big].
\end{align}

In view of the non-renormalisable nature of interacting spin~-~3/2 theories, we can only write a generic set of interactions respecting the SM gauge symmetry between the singlet Dirac-vector spinor, $\chi_{\mu}$, with SM fermions mediated by a vector particle $Z^{\prime}_{\mu}$ as (see for example~\cite{Hassanain:2009at}) 
\begin{align}
\mathcal{L}_{\chi, Z^\prime} + \mathcal{L}_{f, Z^\prime}  &\supset \bar{\chi}_{\alpha} \gamma^{\mu} (g_{\chi}^{V} -
                        \gamma^{5} g_{\chi}^{A}) \chi_{\beta} Z_{\mu}^{\prime} g^{\alpha\beta} \nonumber \\
&+ \sum_{f = q, l, \nu} \bar{f} \gamma^{\mu} ( g_{f}^{V} - \gamma^{5} g_{f}^{A} ) f  Z_{\mu}^{\prime},\label{lag}
\end{align}
where the sum is over all quarks, charged leptons and neutrinos. The interaction is not restricted by MFV to be either a pure vector or axial vector. Although the form of the low energy interactions of spin~-~3/2 particles should arise from an underlying theory at high energies, such as string theory, we follow the approach of simplified model theories. The purpose of a simplified model approach is to characterise the DM production present in a complete theory, without having to specify the complete theory. In these theories the mediator provides the link between the SM and DM candidate.      
In general this interaction will induce flavour-changing neutral currents, which are strongly constrained by 
low energy phenomenology. The constraints can be avoided by imposing a MFV structure on the couplings, 
or by restricting the interactions to one generation.

There exists an extensive range of models with an extra $U(1)^{\prime}$ symmetry (for a review see~\cite{Langacker:2008yv}). The most stringent indirect constraints on $m_{Z^{\prime}}$ arise from the effect of a $Z^{\prime}$ coupling to SM fermions in precision electro-weak observables from low energy weak neutral current experiments~\cite{Han:2013mra,delAguila:2010mx}, and gives a lower limit on $m_{Z^{\prime}}$ of $\mathcal{O}$(1 TeV); where LHC experiments set strong bounds on the $Z^{\prime}$ mass. For a $Z^{\prime}$ coupling with SM particles to be of the order of SM - Z electro-weak coupling this bound is typically $m_Z^{\prime}\geq$ 2 TeV~\cite{Abdallah:2014hon}. This bound is somewhat relaxed (depending on the model) when $Z^{\prime}$ is allowed to decay into DM candidate~\cite{Arcadi:2013qia,Han:2013mra}.  
\begin{figure*}[t]
	\centering
	\includegraphics[height=110pt]{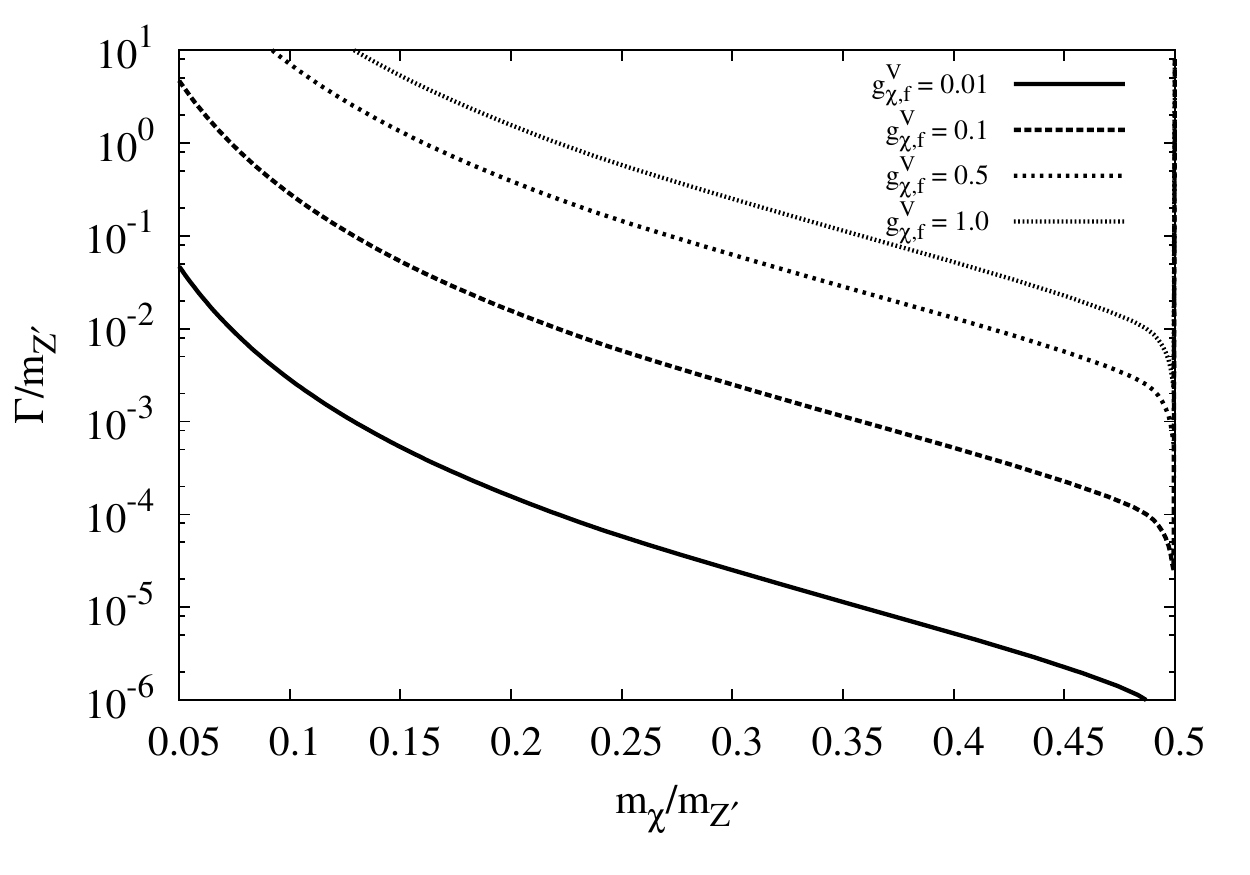}
	\includegraphics[height=110pt]{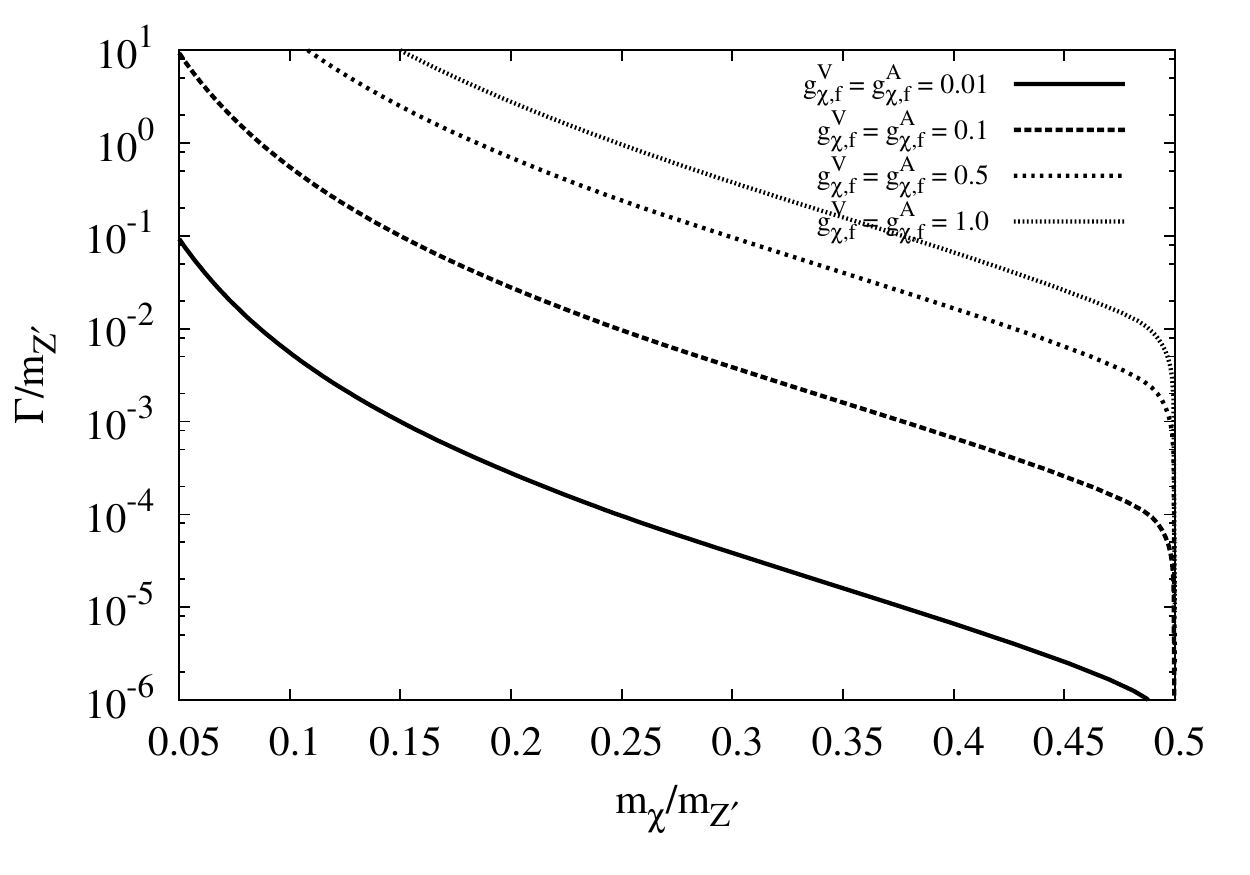}
	\includegraphics[height=110pt]{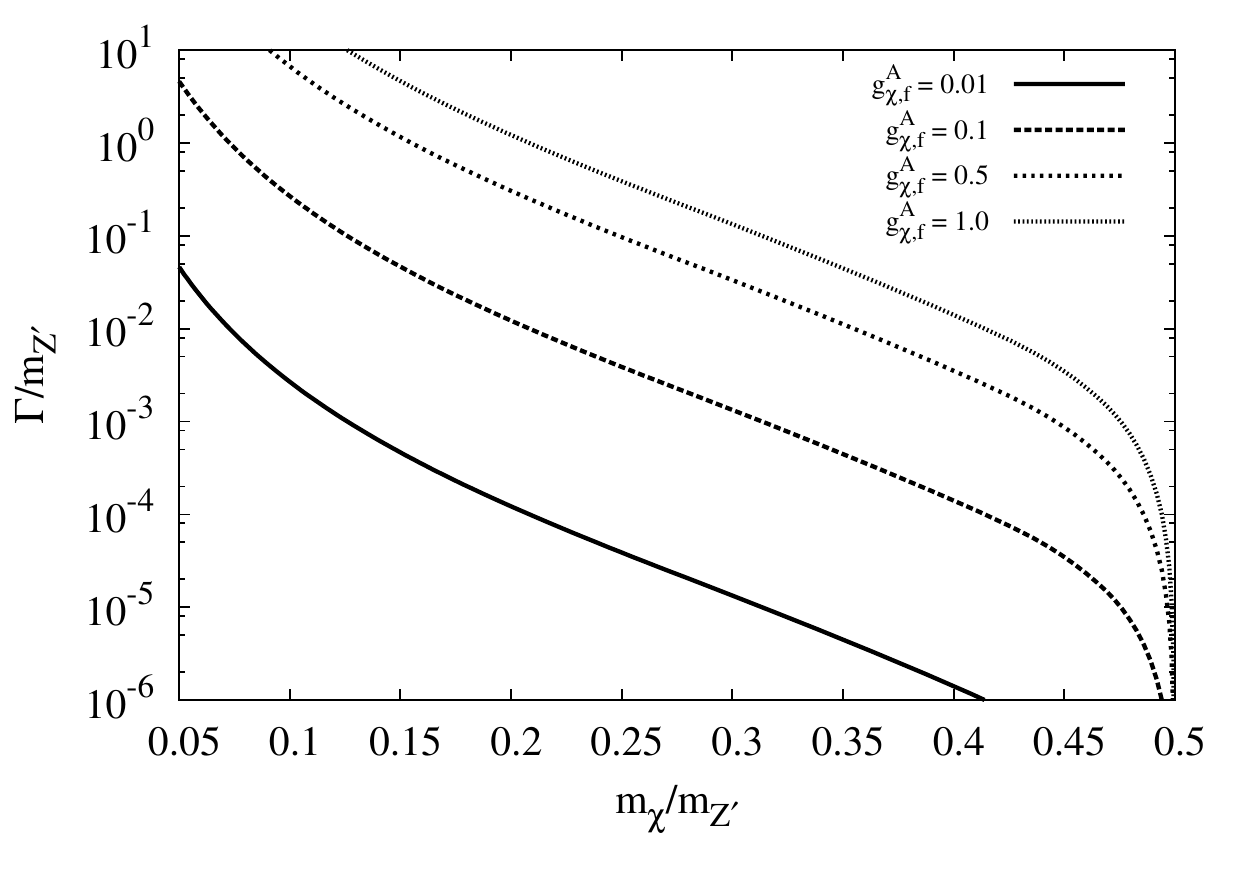}
	\caption{\label{Fig:1} Ratio of the mediator decay width to its mass $\Gamma/{m_{Z^{\prime}}}$ 
	as a functions of $m_{\chi}/{m_{Z^{\prime}}}$ for a few benchmark values of the couplings: 0.1, 0.5 and 1.0. 
        The left panel is for the vector couplings 
	$g^{V}_{\chi,f}$\,, and the middle panel is for the chiral couplings ($g^{V}_{\chi,f} = \pm \,g^{A}_{\chi,f}$).
	The right panel is for the axial-vector couplings $g^{A}_{\chi,f}$.}
\end{figure*}

The decay width $\Gamma(Z^{\prime}\rightarrow f\bar{f} + \chi_{\alpha}\bar{\chi}_{\alpha})$ is given by
\begin{align}
&\Gamma(Z^{\prime}\rightarrow f\bar{f} + \chi_{\alpha}\bar{\chi}_{\alpha}) = 
\sum_{f}\frac{N_{c}}{12\pi}m_{Z^{\prime}}\sqrt{1-\frac{4m_{f}^{2}}{m^{2}_{Z^{\prime}}}}
\nonumber\\
&\quad \quad \times\left[ \left( (g^{V}_{f})^{2} + (g^{A}_{f})^{2} \right) 
 + \frac{2m^{2}_{f}}{m^{2}_{Z^{\prime}}} \left( (g^{V}_{f})^{2} - 2 (g^{A}_{f})^{2}\right) \right] \nonumber\\
&\quad \quad+ \frac{m_{Z^{\prime}}}{108\pi} \left(\frac{m^{2}_{\chi}}{m^{2}_{Z^{\prime}}}\right) 
\sqrt{1-\frac{4m_{\chi}^{2}}{m^{2}_{Z^{\prime}}}} \nonumber \\
&\quad\quad \times \Bigg[ (g_{\chi}^{V})^{2}\left(36 - 2\frac{m^{2}_{Z^{\prime}}}{m^{2}_{\chi}}- 2\frac{m^{4}_{Z^{\prime}}}{m^{4}_{\chi}} 
+ \frac{m^{6}_{Z^{\prime}}}{m^{6}_{\chi}}\right) \nonumber\\
&\qquad\qquad + (g_{\chi}^{A})^{2} \left(-40 + 26\frac{m^{2}_{Z^{\prime}}}{m^{2}_{\chi}} - 8 \frac{m^{4}_{Z^{\prime}}}{m^{4}_{\chi}} 
+ \frac{m^{6}_{Z^{\prime}}}{m^{6}_{\chi}}\right) \Bigg].
\end{align}
The sum extends over all SM fermions $f$ that are above the threshold, $N_{c} = 3$ for quarks and 1 for leptons.

There are several interesting consequences on the DM mass and couplings arising from the above decay width 
expressions. If the DM mass $m_{\chi} > m_{Z^{\prime}}/2$, the only decay channel available to 
the mediator $Z^{\prime}$ is into SM fermions. Since $\Gamma(Z^{\prime}) < \, m_{Z^{\prime}}$ is required for the mediator description to be  perturbatively valid, the vector coupling, for example, should satisfy  
\begin{equation}
\frac{8 m_{Z^{\prime}}}{12\pi}\left(g_{f}^{V}\right)^{2} < m_{Z^{\prime}} \quad \Rightarrow \quad \left(g_{f}^{V}\right)^{2} < \frac{3\pi}{2}.
\end{equation}
Here we consider the coupling to be only to one generation for the purposes of illustration. The qualitative result remains essentially unchanged if all three generations are taken, except that the top quark mass may not be neglected in comparison to the mediator mass. This gives $\Gamma_{Z^{\prime}}/m_{Z^{\prime}} \simeq \frac{2}{3\pi}(g_{f}^{V})^{2}$, and we have the narrow width approximation being applicable for $g_{f}^{V}\leq 1$. However, if the DM mass $m_{\chi} < m_{Z^{\prime}}/2$, the mediator can decay into DM pairs, and there exists a minimum limit on the DM mass $\chi$ for a given value of the mediator mass with the coupling given roughly by    
\begin{equation}
\frac{1}{108\pi}\left(\frac{m_{Z^{\prime}}}{m_{\chi}}\right)^{4} \left(g_{\chi}^{V,A}\right)^{2} < 1. \label{limit}
\end{equation}
If the DM mass is below this value, the decay width would exceed the mediator mass.

In what follows we consider universal couplings for simplicity, $g_{\chi}^{V} = g_{f}^{V}$ and $g_{\chi}^{A} = g_{f}^{A}$, and restrict ourselves to one generation of SM fermions. In Figure~\ref{Fig:1} we have plotted the mediator $Z^{\prime}$ decay width as a function of $m_{\chi}$ for some benchmark values of pure vector couplings $g_{\chi, f}^{V}$, chiral couplings $g_{\chi, f}^{V} = \pm \,g_{\chi, f}^{A}$
and pure axial couplings $g_{\chi, f}^{A}$. It can be seen from Figure~\ref{Fig:1}, that there exists a minimum $m_\chi$ for a given coupling, where a mass of $\chi$ less than the limit given in Equation~\ref{limit}, results the value of decay width more than the value
of $m_\chi$.
This feature is peculiar to the spin~-~3/2 nature of the DM.  

\begin{figure*}
	\centering
	\includegraphics[height=110pt]{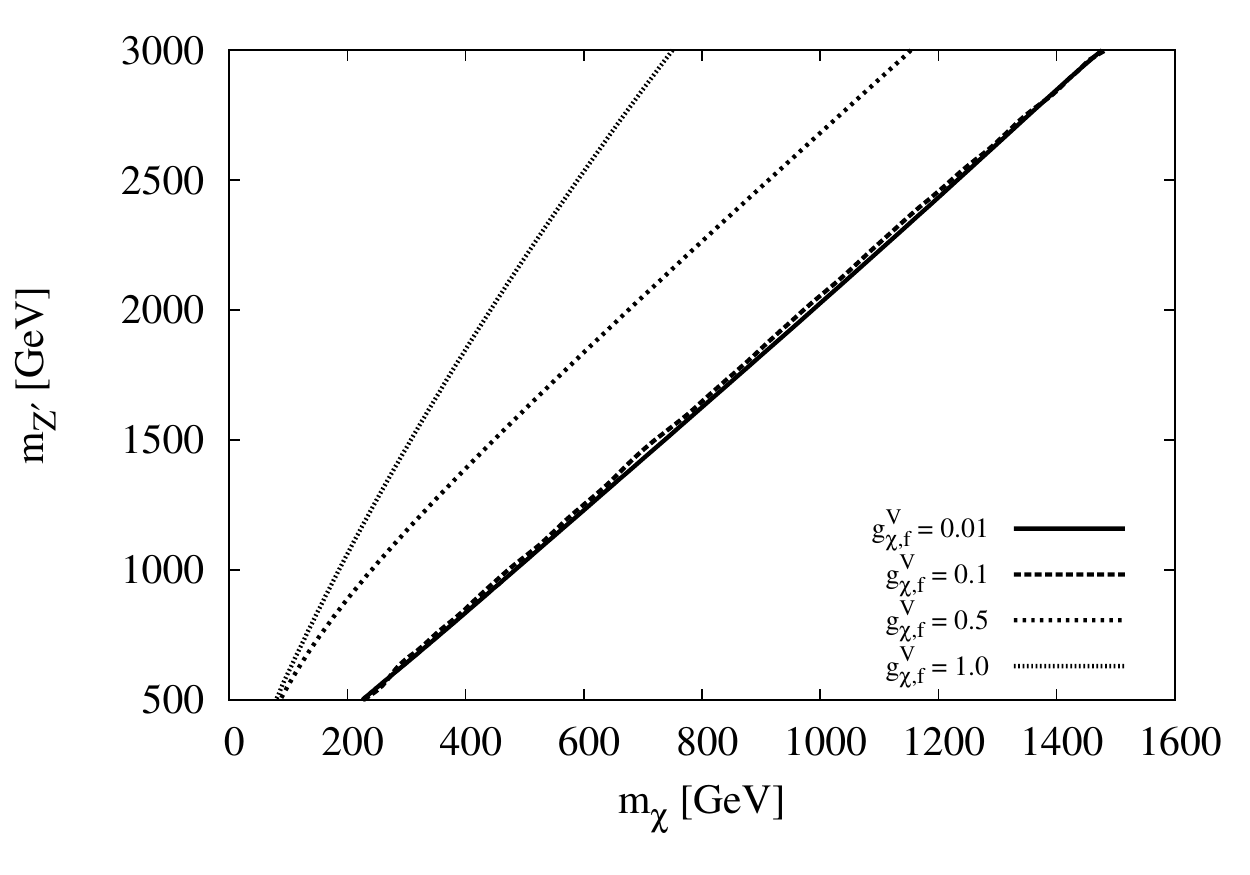}
	\includegraphics[height=110pt]{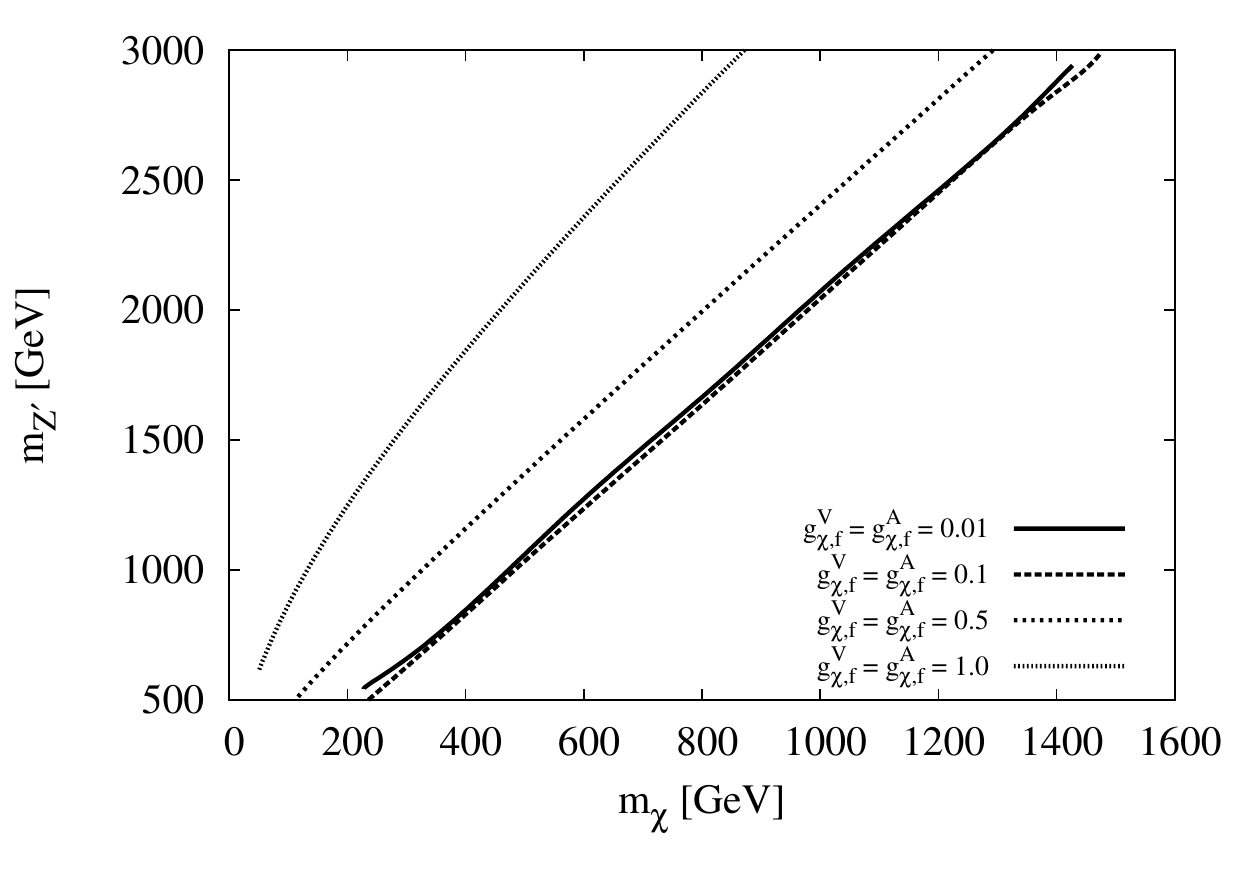}
	\includegraphics[height=110pt]{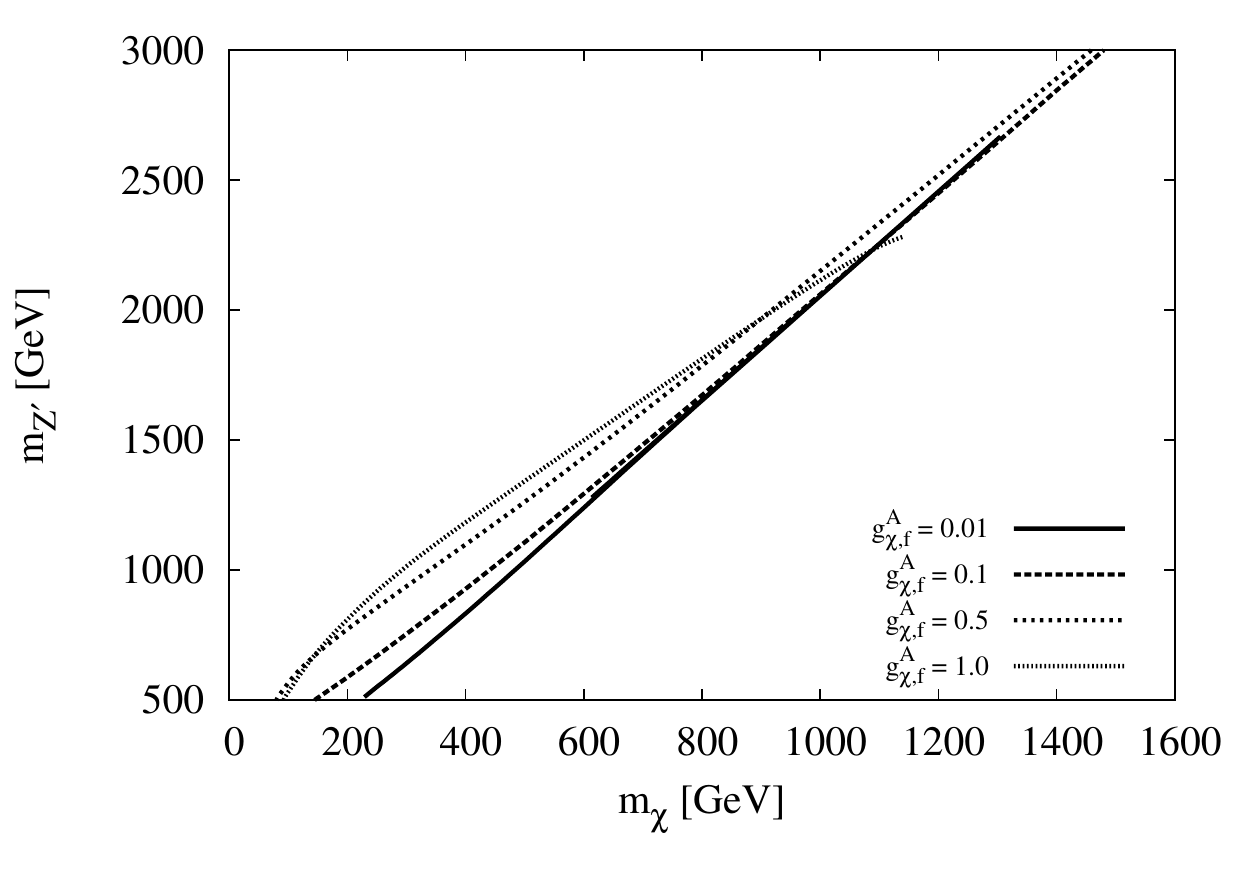}
\caption{\label{Fig:2} The contour plots between the $m_{Z^{\prime}}$ and $m_{\chi}$, where we have assumed that 
the DM $\chi$ saturates the observed DM density. The left and the right panels are for benchmark values of vector and axial-vector couplings respectively. The middle panel is for the chiral coupling.} 
\end{figure*}

\section{Constraints}\label{sec:3}
\subsection{Relic Density}\label{sec:A}
In the early universe the DM particles were kept in thermal equilibrium with the rest of the plasma 
through the creation and annihilation of  $\chi$'s. The cross-section of the annihilation process 
$\chi\bar{\chi} \rightarrow f \bar{f}$ proceeds through $Z^{\prime}$, and the spin averaged cross-section 
is given by
\begin{align}
&\sigma \left(\chi \bar \chi \rightarrow f \bar f \right) = \nonumber \\
&\,  \sum_{f}\frac{N_{c}\,\sqrt{s-4m_{f}^{2}}}{ 432 \,\pi\, m_\chi^4 \,m_{Z^\prime}^4\, s }\frac{1}{\sqrt{s-4m_{\chi}^{2}}} 
 \left[\frac{1}{\left(s - m_{Z^\prime}^2 \right)^2 + \Gamma^2 m_{Z^\prime}^2 }\right]
\nonumber \\ 
&\,\, \times \Big[ \left( g^{A}_{f}\right)^{2} \Big\{ \left( g^{A}_{\chi} \right)^{2} \Big\{ 4m_{f}^{2} \Big\{ 10\, m_{\chi}^{6} 
\left(7m_{Z^{\prime}}^{4}  - 6 m_{Z^{\prime}}^{2}s + 3s^{2} \right)
\nonumber \\ 
&\,\quad  - 2 \, m_{\chi}^{4} \,s \left(16 \,m_{Z^{\prime}}^{4} - 6\,m_{Z^{\prime}}^{2}\,s + 3\,s^{2}\right) 
\nonumber \\ 
&\,\quad + m_{\chi}^{2}\,s^{2}\left(11\,m_{Z^{\prime}}^{4}- 6\,m_{Z^{\prime}}^{2}\,s + 3\,s^{2}\right) - m_{Z^{\prime}}^{4}\,s^{3}\Big\} 
\nonumber \\ 
&\,\quad + m_{Z^{\prime}}^{4} \,s \left(- 40 \,m_{\chi}^{6} + 26\, m_{\chi}^{4} \,s - 8 \,m_{\chi}^{2} \,s^{2}  + s^{3} \right)\Big\}
\nonumber \\
&\,\quad - \left(g^{V}_{\chi}\right)^{2} m_{Z^{\prime}}^{4} \left(4m_{f}^{2}-s\right) \left( 36\,m_{\chi}^{6}  - 2\,m_{\chi}^{4}\,s
- 2\,m_{\chi}^{2}\,s^{2} + s^{3}\right)\Big\}
\nonumber \\
&\,\quad  + \left( g^{V}_{f}\right)^{2} m_{Z^{\prime}}^{4} \left(2 m_{f}^{2} + s \right) 
\nonumber\\  
&\,\qquad \times\Big\{ \left(g_{\chi}^{A}\right)^{2}
\left(-40 \,m_{\chi}^{6} + 26 \,m_{\chi}^{4}s - 8\,m_{\chi}^{2} \,s^{2} + s^{3}\right) 
\nonumber\\ 
&\,\qquad\quad + (g^{V}_{\chi})^{2} \left( 36\,m_{\chi}^{6} - 2\,m_{\chi}^{4}\,s - 2\,m_{\chi}^{2}\,s^{2} + s^{3}\right)\Big\}\Big].
\end{align}
Freeze out occurs when the $\chi$'s are non-relativistic $(v  \ll  c)$. We then have
\begin{equation}
s \simeq 4\,m_{\chi}^{2} + m_{\chi}^{2} v^{2} + \mathcal{O}(v^{4})
\end{equation}
in the lab frame. The cross-section can be expanded in powers of $v^{2}$ as  
\begin{equation}
\sigma v = a + b v^{2} + \mathcal{O}(v^{4}).
\end{equation}
The relic density contributions of the DM particles can be obtained by numerically solving the 
Boltzmann equation:
\begin{equation}
\frac{d n_{\chi}}{d t} + 3 H n_{\chi} = -\langle\sigma |v|\rangle \left(n^{2}_{\chi} - (n^{eq}_{\chi})^{2}\right),
\end{equation}
 where $\langle\sigma |v\rangle$ is the thermally averaged \(\chi\)-annihilation cross-section \(\langle\sigma(\chi\bar{\chi} \rightarrow f\bar{f})|v\rangle\), and $n_{\chi}$ is the number density of the $\chi$'s. When we are in thermal equilibrium the number density is given by
\begin{equation}
n^{eq}_{\chi} = 4\left(\frac{m_{\chi}\, T}{2\pi}\right)^{3/2} \exp\left(-\frac{m_{\chi}}{T}\right).
\end{equation}
The Hubble expansion rate is given by 
\begin{equation}
H = \sqrt{\frac{8\pi \rho}{3M_{\text {pl}}^{2}}},
\end{equation}
where \( M_{\text {pl}} = 1.22\times 10^{19}\) GeV is the Planck mass.
The Boltzmann equation is solved numerically to yield~\cite{Kolb:1990vq}
\begin{equation}
\Omega_{DM} h^{2}\simeq \frac{2\times1.07\times10^{9}X_{F}}{M_{pl}\sqrt{g_{*}}(a + \frac{3b}{X_{F}})},
\end{equation}
where \(g_{*}\) is the number of degrees of freedom at freeze-out temperature \(T_{F}\), and is taken to be \(92\) 
for \(m_{b} < T_{F} < m_{Z^{\prime}}\), \(X_{F} = m_{\chi}/{T_{F}}\). The freeze-out temperature is 
obtained by solving 
\begin{equation}
X_{F} = \ln\Bigg[c(c+2)\sqrt{\frac{45}{8}}\frac{g M_{pl} m_{\chi}(a + \frac{6b}{X_{F}})}{2\pi^{3}\sqrt{g_{*}(X_{F})}\sqrt{X_{F}}}\Bigg],
\end{equation}
where \(c\) is taken to be \(1/2\). For spin~-~3/2 DM \(g = 4\). 

In Figure~\ref{Fig:2} we show the contour graphs between the mass of the mediator \(m_{Z^{\prime}}\) 
and the DM mass \(m_{\chi}\), by assuming that the DM \(\chi\) saturates the observed DM density. From the graphs we see that for small couplings $g \leq  0.1$, the parameter space $(m_{\chi}, m_{Z^{\prime}})$ is consistent with the observed relic density and is thus independent of the coupling. This can be understood by noting that the leading term in the thermally averaged DM annihilation cross-section into SM fermions is given by 
\begin{align}
\langle\sigma(\chi\bar{\chi} \rightarrow f\bar{f})|v\rangle & \simeq \frac{20\, g^{4}}{9\, \pi\, m^{2}_{Z^{\prime}}}\frac{m^{2}_{\chi}}{ m^{2}_{Z^{\prime}}}\frac{1}{\left(1- \frac{4m^{2}_{\chi}}{m^{2}_{Z^{\prime}}}\right)^{2}+ \left(\frac{\Gamma^{2}}{m^{2}_{Z^{\prime}}}\right)}\nonumber\\
& \simeq \frac{8 \times 10^{-24}}{(m_{Z^{\prime}}/1 TeV)^{2}}\frac{g^{4}}{\left(1- \frac{4m^{2}_{\chi}}{m^{2}_{Z^{\prime}}}\right)^{2}+ \left(\frac{\Gamma^{2}}{m^{2}_{Z^{\prime}}}\right)}\nonumber\\
&\times\left(\frac{m_{\chi}}{m_{Z^{\prime}}}\right)^{2} cm^{3}\,s^{-1}.
\end{align}
The annihilation cross-section, being proportional to the fourth power in coupling, falls rapidly for couplings $\leq$ 0.1, and the freeze out occurs early when the temperature is high. This will result in the relic density falling below the observed value. The annihilation rate, however, receives resonant enhancement at $m_{\chi} \simeq \frac{1}{2} m_{Z^{\prime}}$, in which case the $\Gamma/m_{Z^{\prime}}$ term dominates over the pole term in the denominator. Thus near resonance the annihilation cross-section becomes
independent of the coupling and we get the relic density contour curves almost independent of coupling. In this situation the observed
relic density is obtained for $m_{\chi} \simeq \frac{1}{2} m_{Z^{\prime}}$ as is evident from the graphs.   
\begin{figure*}
	\centering
	\includegraphics[height=150pt]{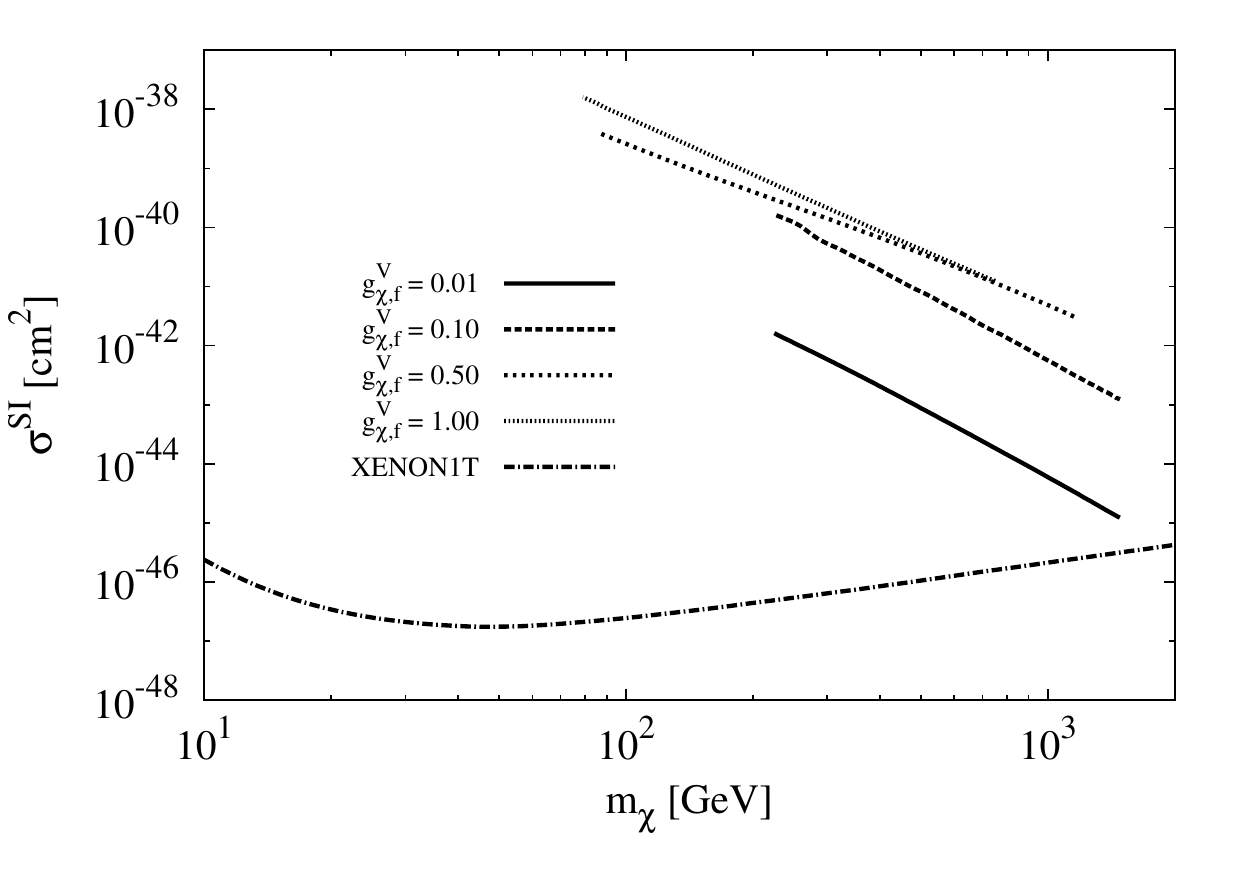}
	\includegraphics[height=150pt]{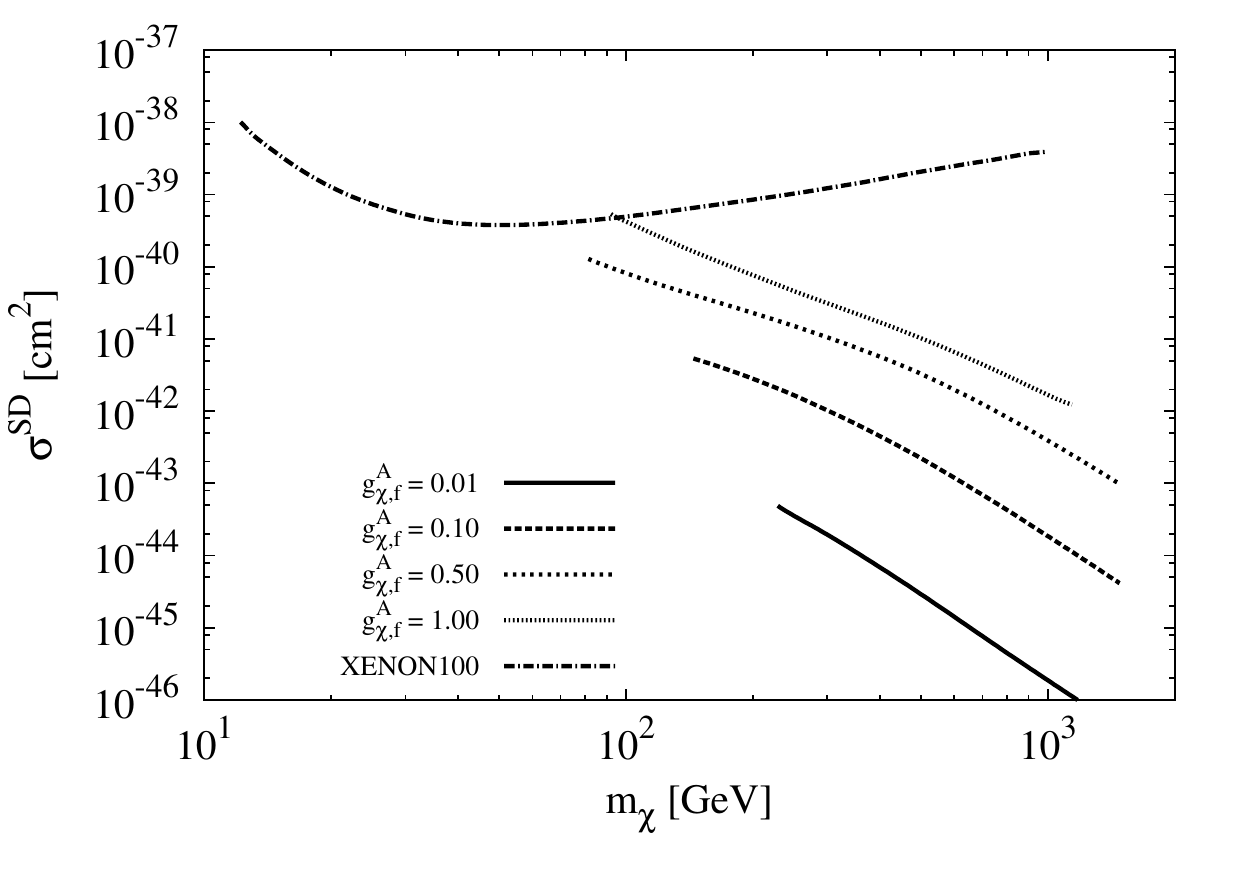}
    \caption{\label{Fig:3} The spin-independent nucleon-DM cross-section $\sigma^{\rm SI}$  (left panel) 
    and spin-dependent nucleon-DM cross-section $\sigma^{\rm SD}$ (right panel). The predicted 
    cross-section is shown here for different values of the coupling, and are in agreement with the 
    relic density constraints. In the plots we show the recent XENON1T data for $\sigma^{\rm SI}$, 
    and the XENON 100 neutron bounds for $\sigma^{\rm SD}$. For the vector coupling almost the entire 
    parameter space $(m_{\chi}, m_{Z^{\prime}})$ is consistent with the relic-density constraints and is ruled 
    out from the XENON1T bound. In contrast for the axial-vector coupling, the parameter space  
    consistent with the observed relic density, is also allowed by the direct XENON 100 neutron bound.}
\end{figure*}

\subsection{Direct Detection}
\label{sec:B}
Constraints from DM detection experiments can be obtained from the elastic DM-nucleon cross-section. 
In the present case, owing to the presence of both vector and axial-vector couplings, the DM-nucleon 
scattering has both spin-independent and spin-dependent components. The corresponding cross-section 
at zero momentum transfer can be easily computed~\cite{DelNobile:2013sia,Freytsis:2010ne,Barger:2008qd}. 
The spin-independent and spin-dependent sub-dominant cross-sections are given by~\cite{Abdallah:2015ter}   
\begin{align}
\sigma_{\chi N}^{\rm SI} &= \frac{\mu^{2} f^{2}_{N}}{\pi m^{4}_{Z^{\prime}}} 
= \frac{9\mu^{2}\left(g_{f}^{V} g_{\chi}^{V}\right)^{2}}{\pi m^{4}_{Z^{\prime}}}
\nonumber\\ 
&\simeq 1.4\times 10^{-37} \left(g_{f}^{V} g_{\chi}^{V}\right)^{2} 
\left(\frac{\mu}{1 \rm GeV}\right)^{2} \left(\frac{300 \,\rm GeV}{m_{Z^{\prime}}}\right)^{4} {\rm cm}^{2},
\end{align}
and 
\begin{align}
\sigma_{\chi N}^{\rm SD} &= \frac{5\mu^{2}}{3\pi m^{4}_{Z^{\prime}}}a^{2}_{N} = \frac{5\mu^{2}\left(g_{f}^{A} 
	g_{\chi}^{A} \right)^{2}}{3\pi m^{4}_{Z^{\prime}}} \left( \Delta u^{N} + \Delta d^{N} + \Delta s^{N}\right)^{2}
\nonumber\\
&\simeq 4.7\times 10^{-39}\left( g_{\chi}^{A} g_{f}^{A}\right)^{2} \left(\frac{\mu}{1 \rm GeV}\right)^{2}
\left(\frac{300 \,\rm GeV}{m_{Z^{\prime}}}\right)^{4} {\rm cm}^{2},
\end{align}
where  
\begin{equation}
\mu = \frac{m_{\chi}m_{N}}{m_{\chi} + m_{N}}
\end{equation}
is the reduced mass. $m_{N} = (m_{p}+m_{n})/2\simeq 0.939$ GeV is the  nucleon-mass for direct detection, 
with $f_{p}$, $f_{n}$ and $a_{p,n}$ being given by:
\begin{equation}
f_{p} = g_{\chi}^{V} \left(2 g_{u}^{V} + g_{d}^{V}\right),\quad\quad\quad f_{n} = g_{\chi}^{V}\left(2g_{d}^{V} + g_{u}^{V}\right) 
\end{equation}
and 
\begin{equation}
a_{p,n} = \sum_{q= u,d,s} g_{\chi}^{A} \Delta q^{p,n} g_{q}^{A}.
\end{equation}
The coefficients $\Delta q^{p,n}$ depend on the light quark contributions to the nucleon spin~\cite{Abdallah:2015ter};
\begin{align}
\Delta u^{p} =& \Delta d^{n} = 0.84\pm 0.02,\nonumber\\
\Delta d^{p} =& \Delta u^{n} = -0.43\pm 0.02, \\
\Delta s^{p} =& \Delta s^{n} = -0.09\pm 0.02. \nonumber
\end{align}
The axial-vector term is suppressed by the momentum transfer, or by the DM velocity, and has been neglected. 
In Figure~\ref{Fig:3} we show the predictions for the spin-independent \(\sigma^{\rm SI}\) and 
spin-dependent \(\sigma^{\rm SD}\) cross-sections for benchmark values of the vector and axial-vector 
couplings, as a function of DM mass $m_{\chi}$. The corresponding experimental bounds from 
XENON1T~\cite{Aprile:2015uzo} and XENON100~\cite{Aprile:2013doa} are also displayed. The mediator mass 
$m_{Z^{\prime}}$ is set to give the observed relic density for all values of $m_{\chi}$ and the couplings.
We find that for the vector coupling almost the entire parameter space $(m_{\chi}, m_{Z^{\prime}})$ consistent 
with the observed relic density, is ruled out from the XENON1T bound on spin-independent nucleon-DM elastic 
scattering cross-sections. The XENON-100 data on the spin-dependent cross-section on the other hand does not 
place severe constraints on the parameter space, and as such the allowed parameter space is consistent with the 
observe relic density. The same is true for the chiral couplings.  
\begin{figure*}
	\centering
	\includegraphics[height=150pt]{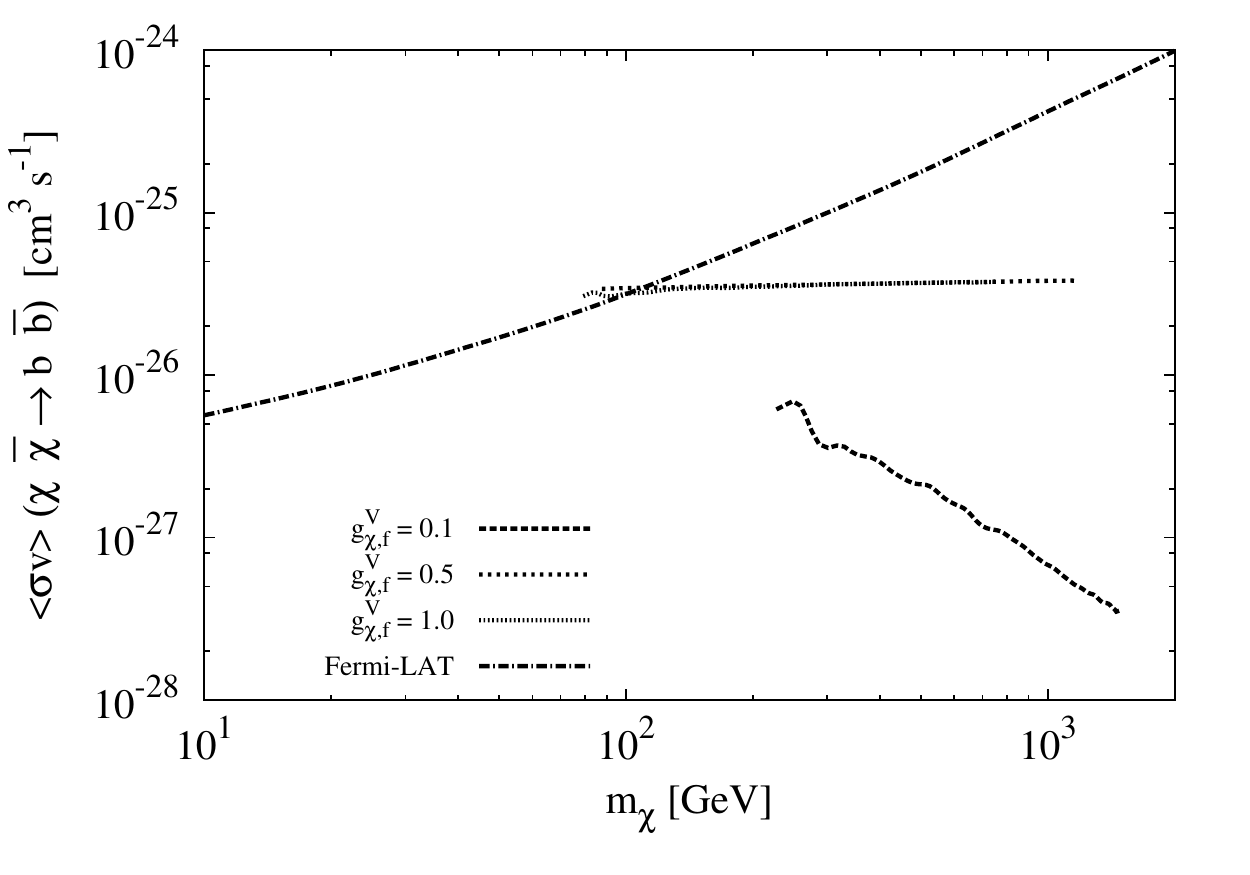}
	\includegraphics[height=150pt]{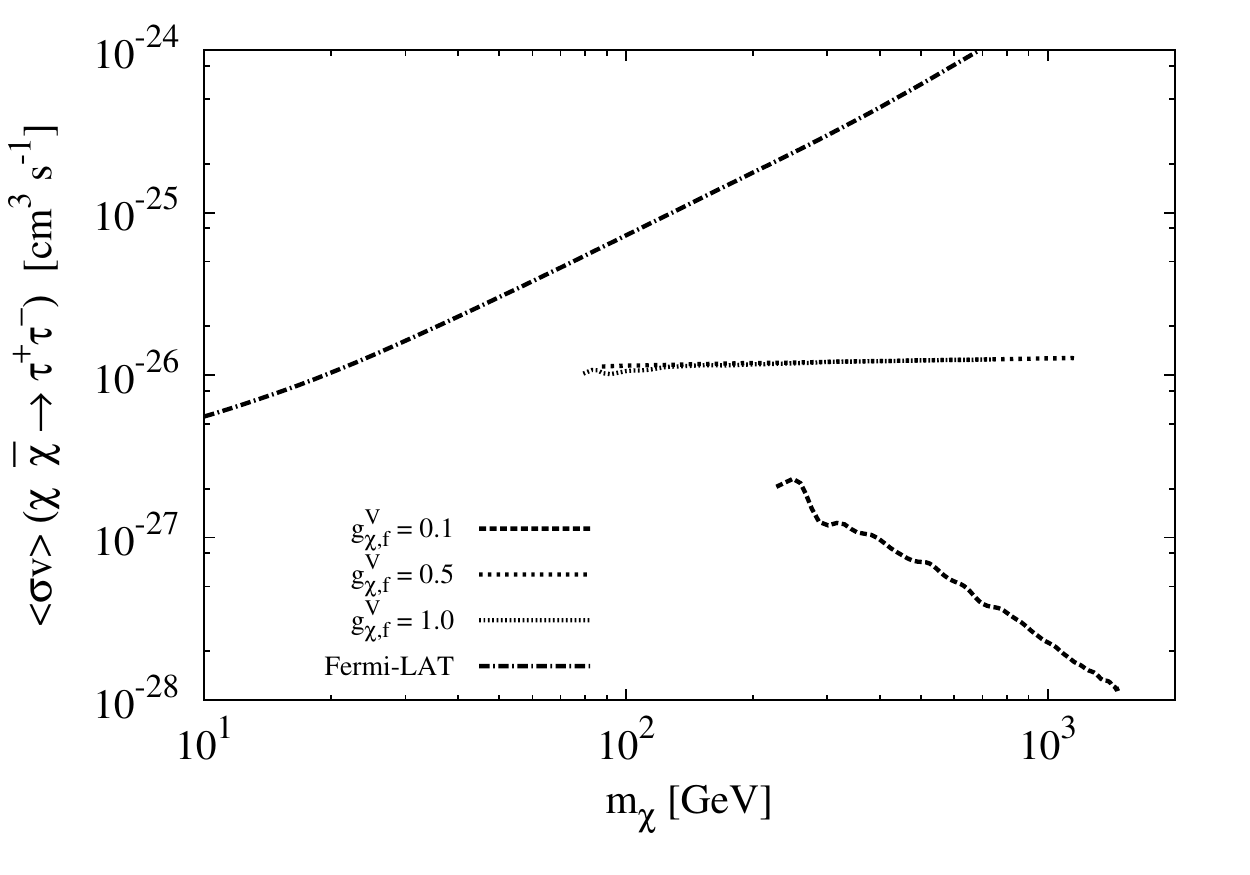}
	\includegraphics[height=150pt]{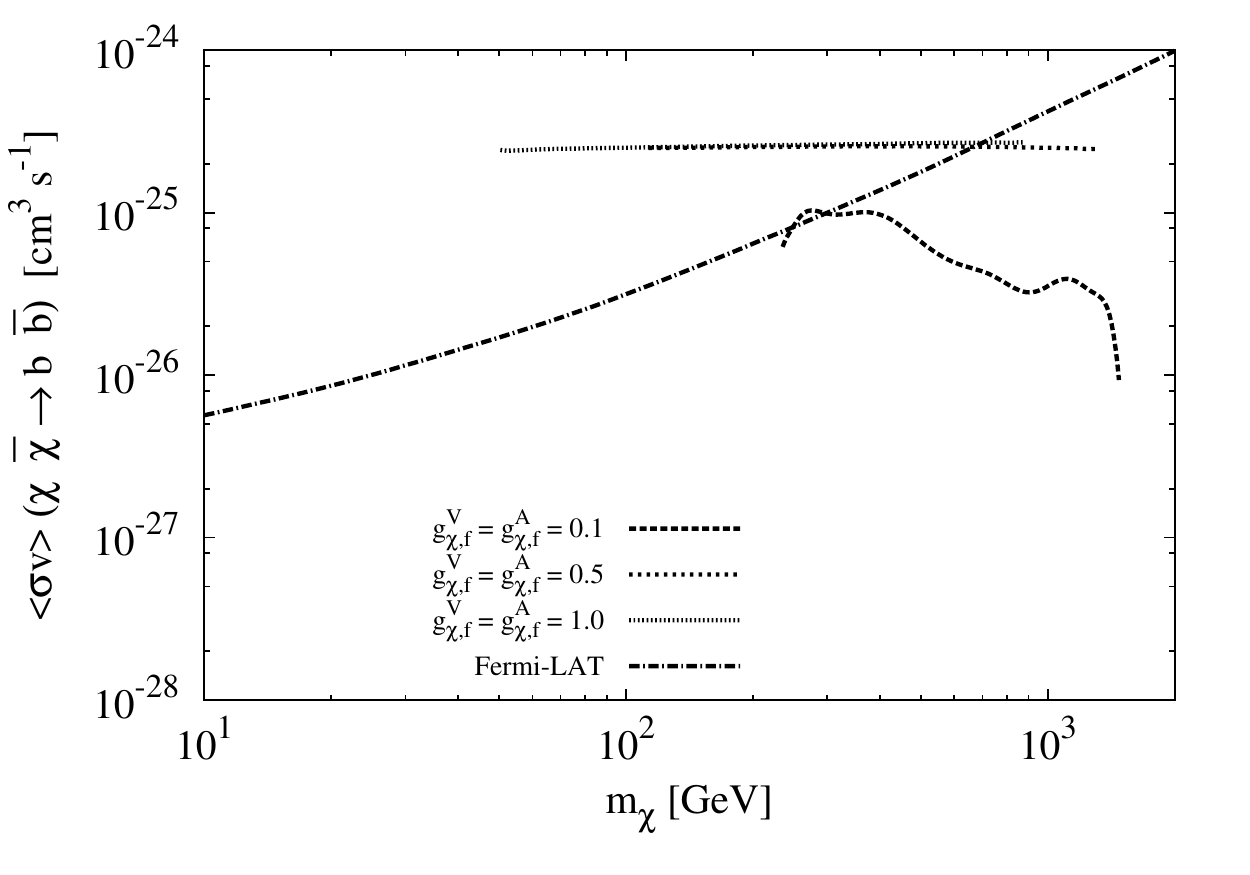}
	\includegraphics[height=150pt]{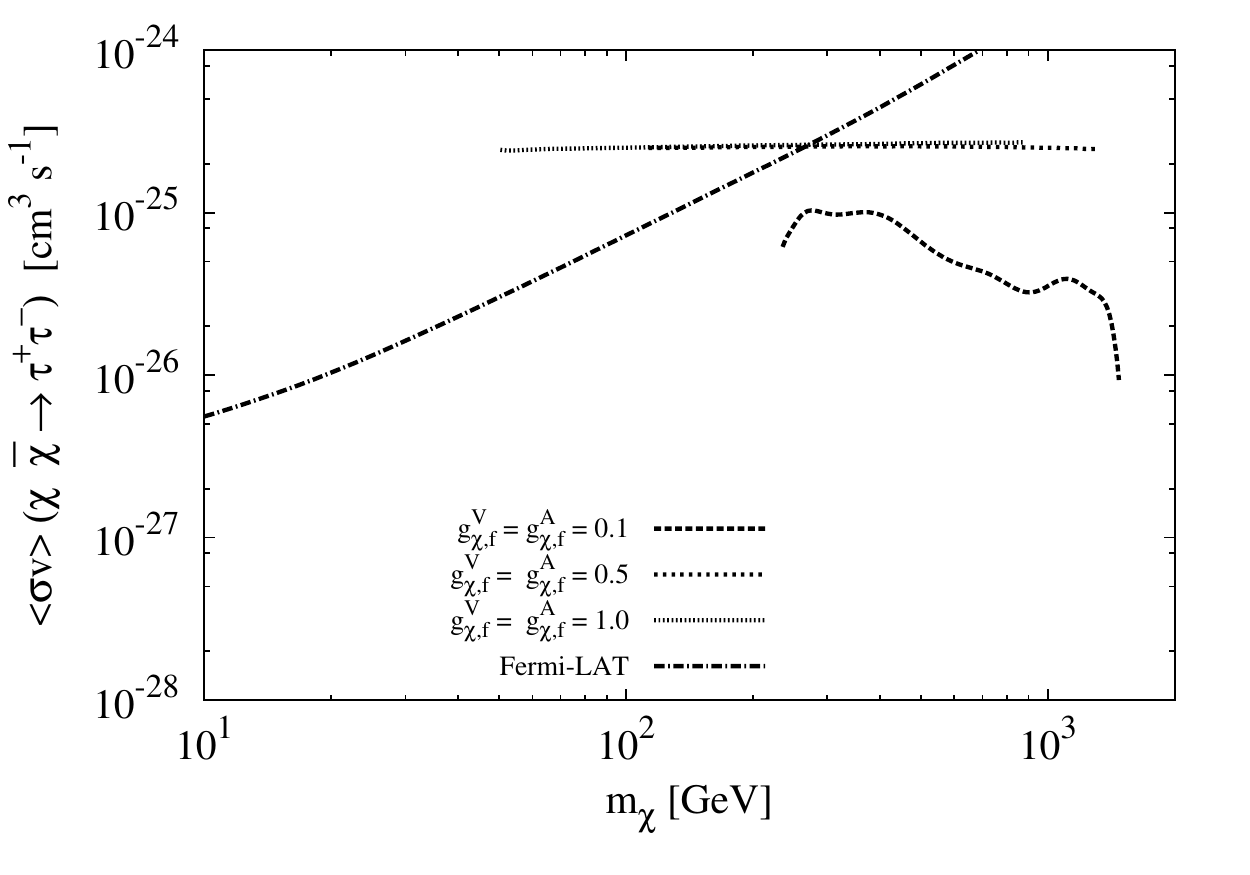}
	\includegraphics[height=150pt]{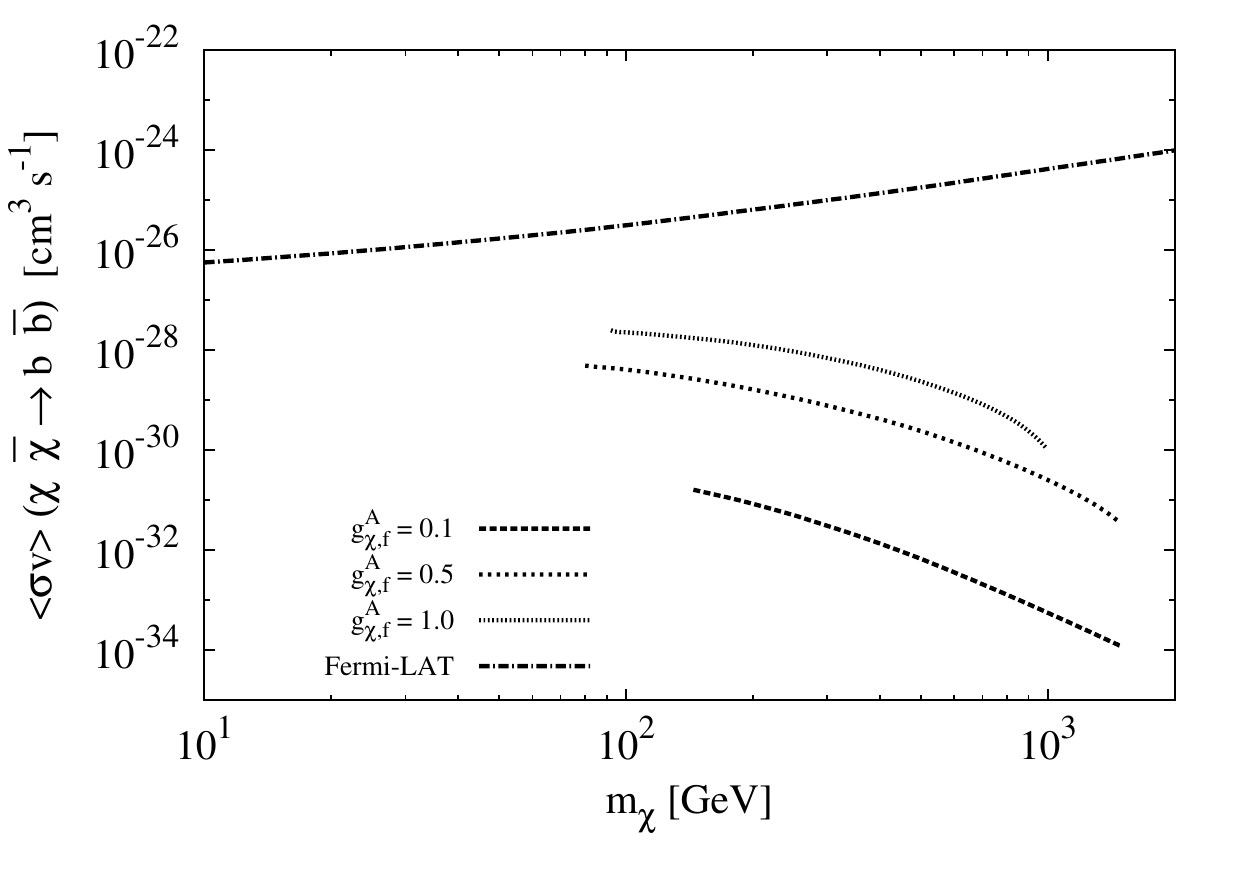}
	\includegraphics[height=150pt]{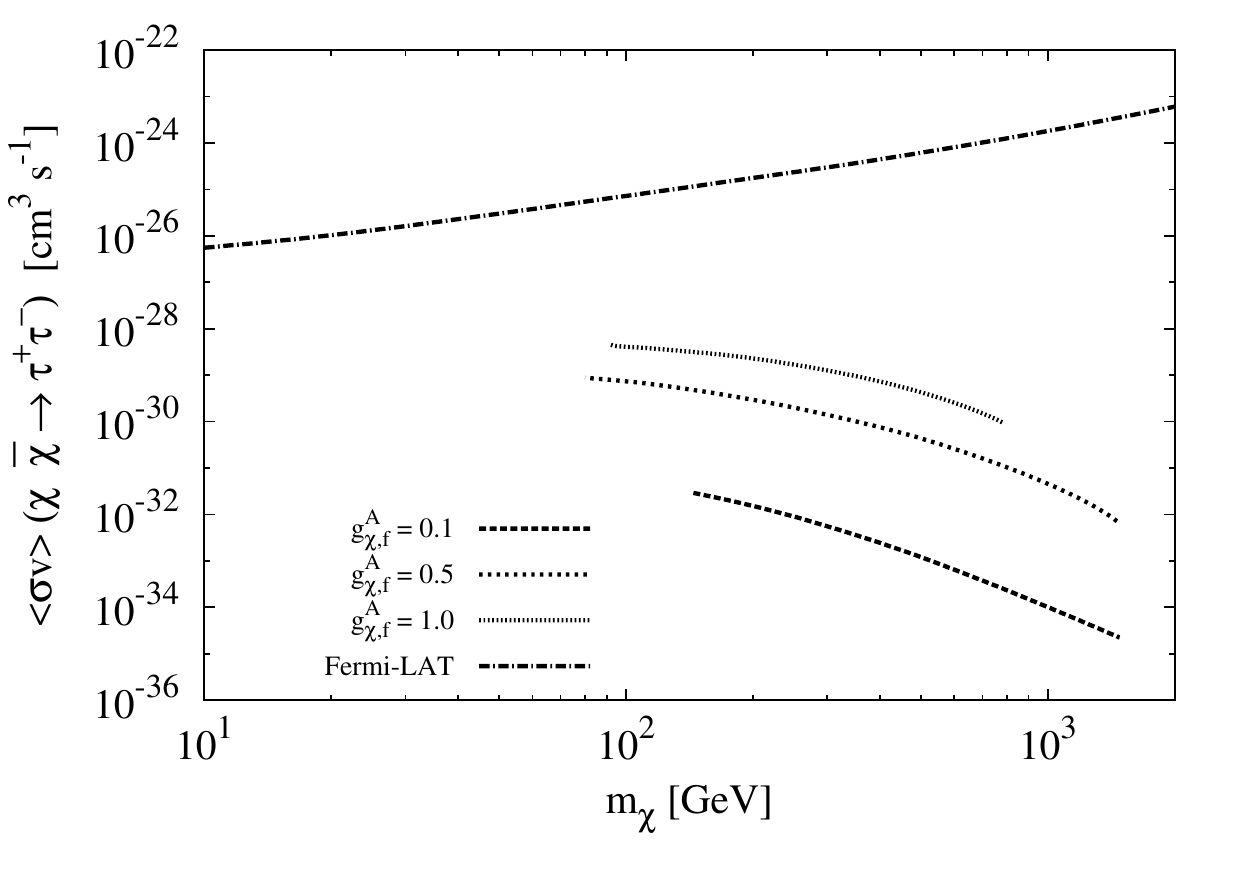}
\caption{\label{Fig:4} The prediction for the DM $\chi$ annihilation rate into $b\bar{b}$ and $\tau^{+}\tau^{-}$ for benchmark values of couplings. The top, middle and bottom panels are for pure vector, chiral and axial couplings respectively. The cross-sections are obtained for $(m_{\chi}, m_{Z^{\prime}})$ values consistent with the observed relic density. 
 Bounds from the Fermi-LAT experiments are also shown.}
\end{figure*}

\subsection{Indirect Detection}
\label{sec:C}
DM annihilation in the universe would result in cosmic ray fluxes which can be observed by dedicated 
detectors. The Fermi Large Area Telescope (LAT) 
collaboration~\cite{Drlica-Wagner:2015xua,Ackermann:2015zua} has produced constraints on 
the DM annihilation cross-section into some final states, namely 
\(e^{+}e^{-}\), \(\mu^{+}\mu^{-}\), \(\tau^{+}\tau^{-}\), \(b\bar{b}\), \(u\bar{u}\), \(W^{+}W^{-}\) 
etc.~\cite{Drlica-Wagner:2015xua,Cherry:2015oca}.

In Figure~\ref{Fig:4} we show the prediction for the DM annihilation into $b\bar{b}$ and $\tau^+ \tau^-$ for vector, axial-vector and chiral  couplings, as a function of $m_{\chi}$. The predictions shown here are for benchmark values of couplings and for the DM mass $m_{\chi}$  consistent with the observed relic density. We have also shown the bounds from the Fermi-LAT experiments. It can be seen from these figures that the Fermi-LAT data on the DM annihilation cross-section, \(\langle\sigma(\chi\bar{\chi} \rightarrow b\bar{b}, \tau^{+}\tau^{-})|v\rangle\), is consistent with the benchmark vector and axial-vector couplings, and for $(m_{\chi}, m_{Z^{\prime}})$ parameters obtained from the observed relic density. However, for the chiral couplings considered in this work there is only a narrow window in the high DM mass ($m_{\chi}  \geq $ 400 GeV) range for the coupling $g \simeq 1$. For small values of the coupling ($g  \leq 0.1$) Fermi-LAT data does not provide any stringent bounds on the $(m_{\chi}, m_{Z^{\prime}})$ parameter space. 
\begin{figure*}
	\centering
	\includegraphics[height=110pt]{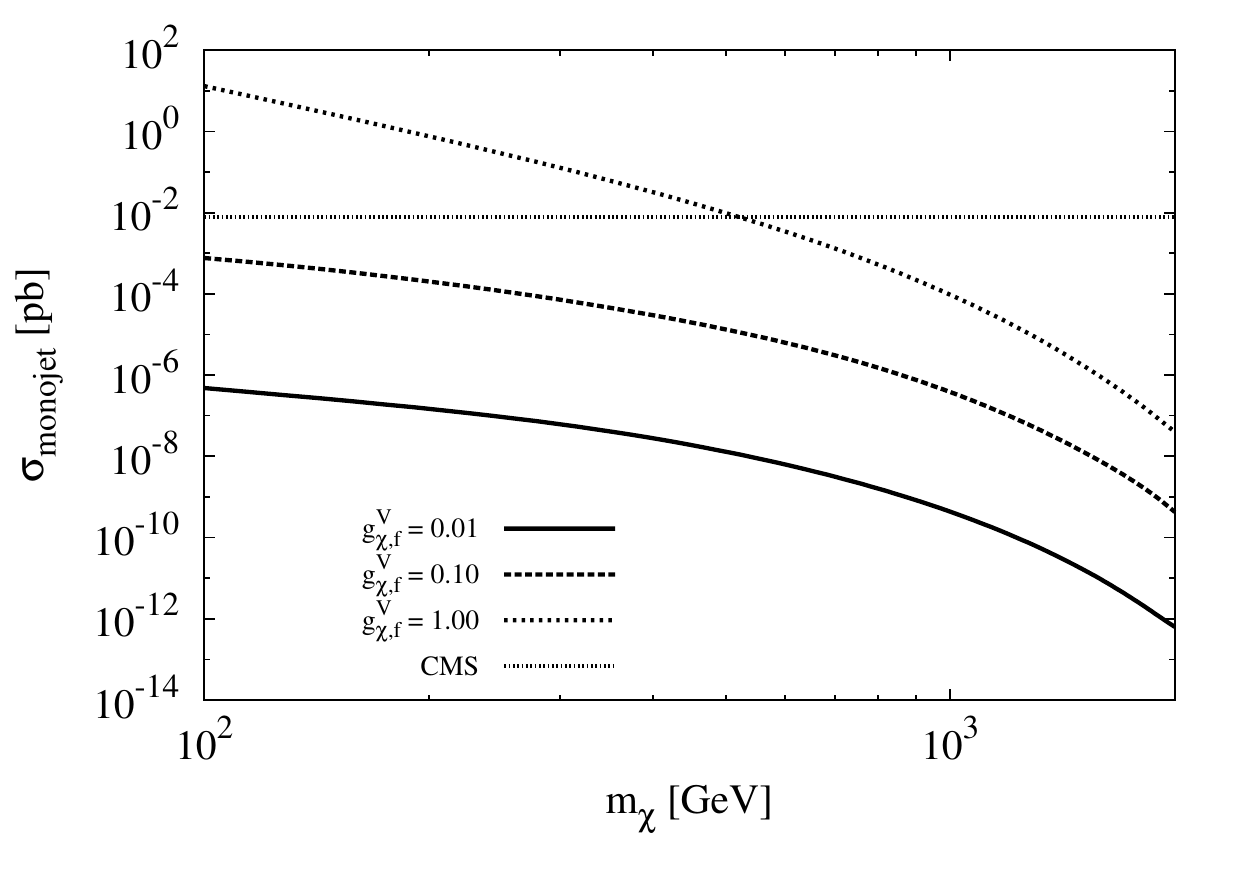}
	\includegraphics[height=110pt]{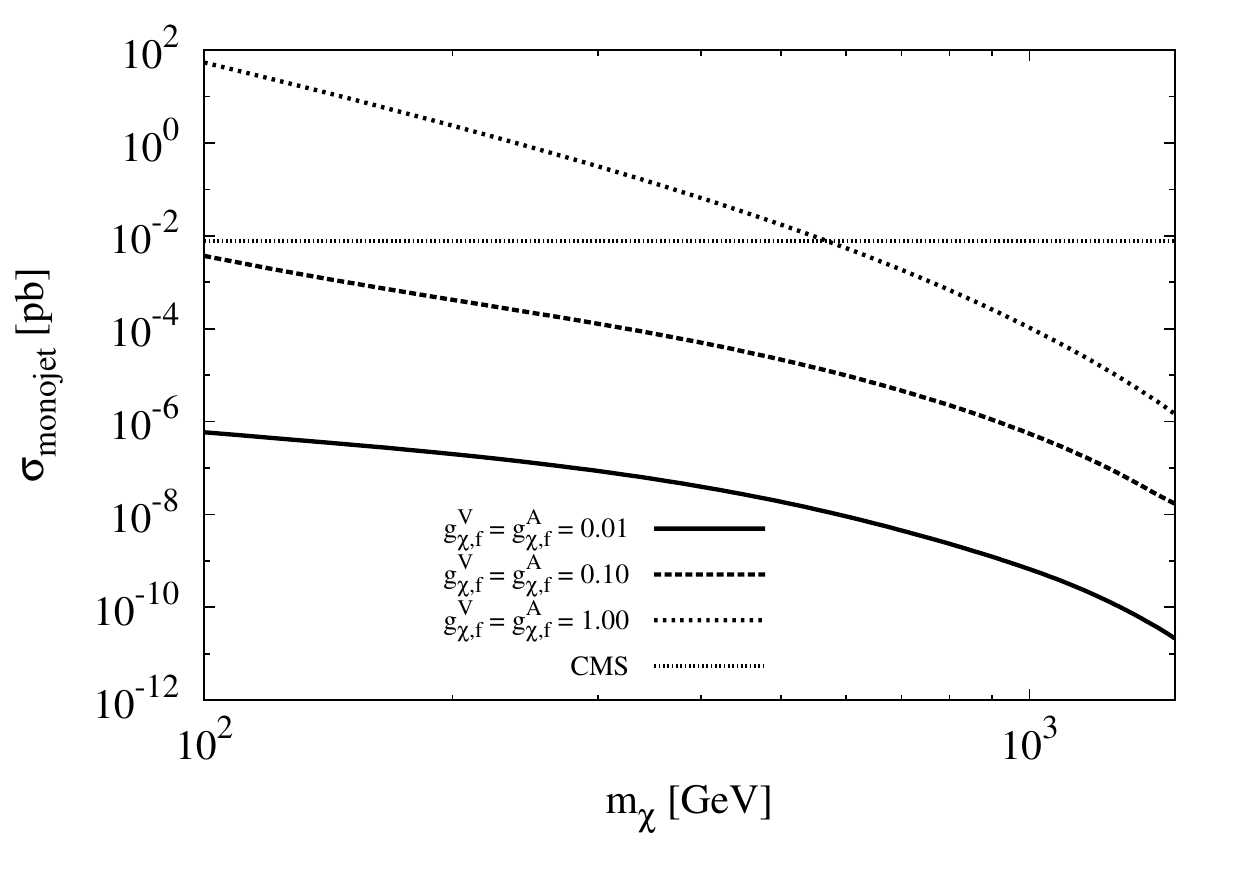}
	\includegraphics[height=110pt]{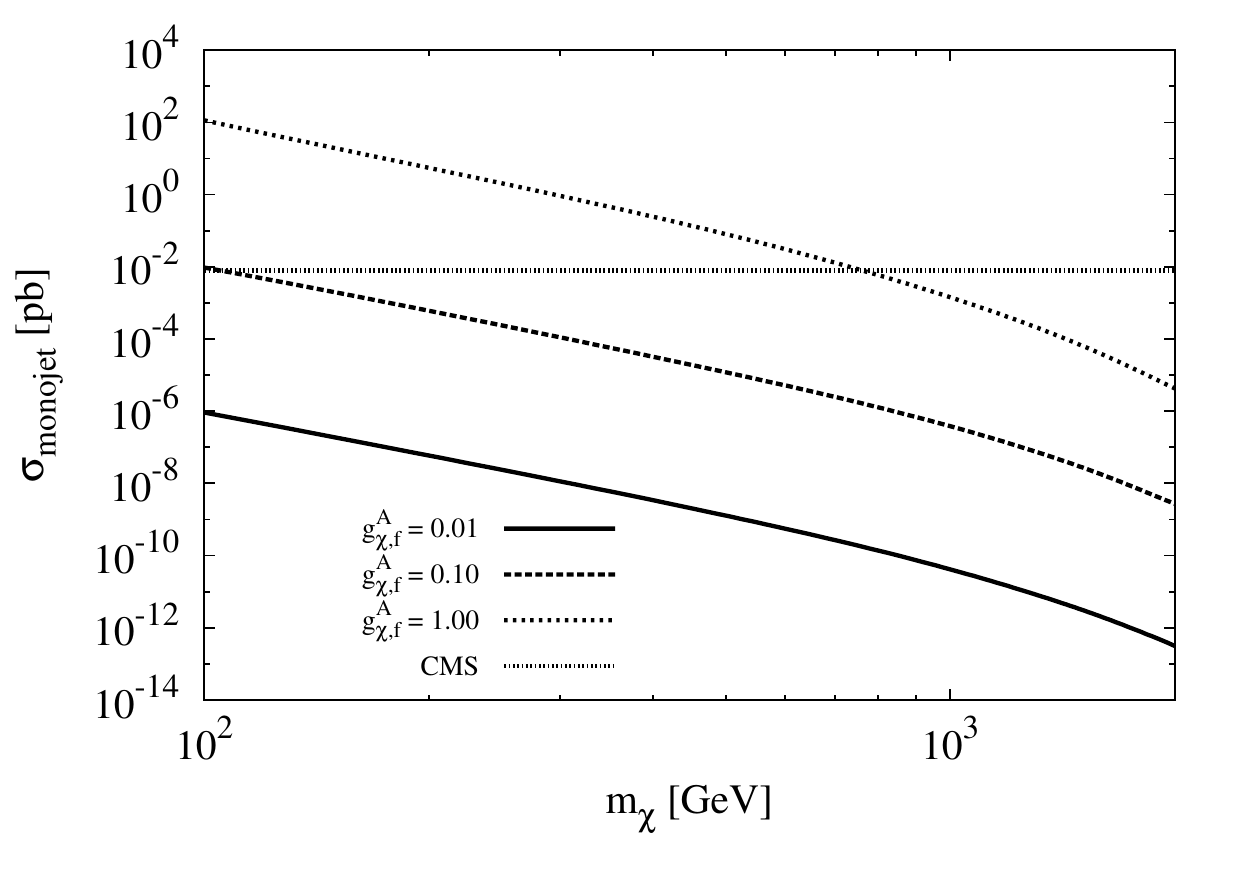}
	\caption{\label{Fig:5} The monojet cross-section in [pb] at the LHC with missing transverse energy 
	 \({\not}E_{T}\) + 1 jet signal, through $pp \to Z^{\prime} \to \chi\bar{\chi}+ 1 j$. The cross-sections 
	 are obtained by considering values of $(m_{Z^{\prime}}, m_\chi)$ consistent with the observed 
	 relic density for the benchmark couplings. The allowed
	 parameter space for spin~-~3/2 DM candidates lies below the CMS bound of $\sigma_{\rm monojet} = 7.8$
	 {\rm fb}, as explained in the text. The left, central and middle panels are for pure vector, chiral and axial-vector couplings respectively.}
\end{figure*}
\subsection{Collider Constraints}
\label{monojet}
Monojet searches at the LHC with missing transverse energy, \({\not}E_{T}\), have been used by CMS 
at 8 TeV, based on an integrated luminosity of 19.7 $\rm fb^{-1}$~\cite{Khachatryan:2014rra}, to put 
constraints on the interaction of quarks and DM particles. In the context of a spin-1/2 DM particle interacting 
through a vector mediator, with vector and axial-vector couplings, constraints on the DM mass $m_{\chi}$ 
and the mediator mass $m_{Z^{\prime}}$ for some representative values of the coupling have been obtained 
in the literature~\cite{Buchmueller:2014yoa, Harris:2014hga, Jacques:2015zha, Chala:2015ama, Alves:2013tqa}. 

For monojet constraints at the LHC, we use the parameter space $(m_{\chi}, m_{Z^{\prime}})$ for the spin~-~3/2 DM, 
consistent with the observed DM density for benchmark couplings. To obtain the cross-section for monojets 
we generate parton level events of the process $p p \to \chi \bar\chi + 1 j$ using {\texttt{MadGraph5}}~\cite{Alwall:2014hca}, 
where the required model file for the Lagrangian~(\ref{lag}) is obtained from {\texttt{FeynRules}}~\cite{Alloul:2013bka}. 
The cross-sections are calculated here to obtain bounds by requiring ${\not}E_{T} > 450$ GeV, for which the 
CMS results exclude new contributions to the monojet cross-section exceeding 7.8 $\rm fb$ at 95$\%$ CL. 
The resulting monojet cross-section for the vector, axial-vector and chiral couplings are shown in Figure~\ref{Fig:5}, 
where we find that the vector coupling results are in agreement with the bounds from the direct detection 
experiments. In the case of axial-vector couplings, the monojet search places stronger constraints on the 
parameters, in comparison to the constraints from direct and indirect searches, albeit for $g^{A}_{\chi, f} \sim 1$.         

\section{Summary}\label{sec:4}
In this paper we have considered a spin~-~3/2 DM particle interacting with SM fermions through a vector mediator 
in the $s$-channel. Assuming MFV we used universal vector and axial-vector couplings and restricted ourselves to one 
generation. The main observations are the following:   
\begin{itemize}
\item In view of the spin~-~3/2 nature of the DM, in addition to the restriction on the coupling arising from the decay 
width, there also exists a minimum value of the DM mass for a given coupling and mediator mass.
\item For the case of vector and chiral couplings, almost the entire parameter space $(m_{\chi}, m_{Z^{\prime}})$ consistent with the 
observed relic density, is ruled out by direct detection through nucleon-DM elastic scattering bounds given by 
XENON1T data. 
\item The case of a vector mediator with pure axial-vector coupling is, in contrast, different with respect to the vector 
coupling. The parameter space is consistent with the observed relic density and is also allowed by the indirect 
and direct (XENON100 neutron) observations.
\item For the benchmark couplings considered here there are no strong bounds on vector and chiral couplings from the 
monojet searches at the LHC, and the results are in broad agreement with the direct detection experiments.
\item The case of pure axial coupling is, however, different. Here the monojet search place stronger constraints 
on the parameters in comparison to the constraints obtained from the XENON100 neutron observations.
\item The Fermi-LAT data on the DM annihilation cross-section is consistent with the vector and axial-vector couplings considered here, and for the $(m_{\chi}, m_{Z^{\prime}})$ parameter values obtained from the relic density. For couplings $g  \leq  0.1$ the Fermi-LAT data does not provide stringent bounds on the $(m_{\chi}, m_{Z^{\prime}})$ parameters. For chiral couplings the data allows only a narrow window in the DM mass ($m_{\chi} \geq\,$400 \,GeV) and $g \simeq 1$.
\item In the EFT frame work for pure vector couplings~\cite{Ding:2012sm,Yu:2011by} the entire parameter space 10 GeV $< m_{\chi} <$ 1 TeV, and an effective interaction scale of the order of a few tens of TeV, though consistent with the observed relic density, is ruled out from the direct detection observations. For the case of pure axial coupling, bounds from direct detection do not forbid the DM mass lying in this range. This is in agreement with our study in a simple $s$-channel mediator model, except that in the mediator model the minimum allowed DM mass is consistent with the observed relic density, and is of order of 100 GeV. In the case of couplings with chiral SM fermions $(g_{f}^{V} = g_{f}^{A})$ it was found~\cite{Ding:2013nvx} that for a spin~-~3/2 DM mass up to 1 TeV, the entire parameter space is ruled out from direct detection. The monojet +  \({\not}E_{T}\) searches at ATLAS rules out DM masses up to 200 GeV. In contrast the $s$-channel mediator model monojet searches at ATLAS are more stringent, and the allowed DM mass limit is raised to greater than 500 GeV. For DM masses exceeding 1 TeV, there are no direct detection constraints, but collider and indirect observation constraints still exist.                
\end{itemize}

\begin{acknowledgements}
MOK is supported by the National Research Foundation of South Africa. ASC is supported in part by the National 
Research Foundation of South Africa (Grant No:91549). AG would like to thank the Mandelstam Institute for 
Theoretical Physics (University of the Witwatersrand) for their support and hospitality during his visit.
\end{acknowledgements}

\end{document}